\documentclass[twocolumn]{raa}
\usepackage{graphicx,times}
\usepackage{natbib}
\usepackage{amssymb,amsmath}
\bibpunct{(}{)}{;}{a}{}{,}

\usepackage[a4paper=true,driverfallback=dvipdfm,pagebackref=true]{hyperref}

\usepackage{enumitem}
\newcommand{\Om}              {\Omega_{\rm m}}
\newcommand{\Ob}              {\Omega_{\rm b}}
\newcommand{\Ol}              {\Omega_{\rm \lambda}}
\newcommand{\seight}          {\sigma_{\rm 8}}
\newcommand{\Hzero}           {H_{\rm 0}}

\newcommand{\Mstar}           {M_{\rm *}}
\newcommand{\Mstarp}          {M_{\rm *}^{\rm 1}}
\newcommand{\Mstars}          {M_{\rm *}^{\rm 2}}
\newcommand{\fgaspair}       {f_{\rm gas}^{\rm pair}}
\newcommand{\Mgas}           {M_{\rm gas}}
\newcommand{\Mgasp}          {M_{\rm gas}^{\rm 1}}
\newcommand{\Mgass}          {M_{\rm gas}^{\rm 2}}
\newcommand{\Rm}             {R_{\rm m}}
\newcommand{\fmerg}          {f_{\rm merg}}
\newcommand{\fgas}           {f_{\rm gas}}
\newcommand{\fmass}          {f_{\rm mass}}
\newcommand{\fmN}            {fm_{\rm N}}
\newcommand{\Mboth}          {M_{\rm *}^{\rm both}}
\newcommand{\fDAGN}          {f_{\rm DAGN}}
\newcommand{\Mcrit}          {M_{\rm bh,crit}}
\newcommand{\Mbh}            {M_{\rm bh}}
\newcommand{\Mtwoh}          {M_{\rm 200}}

\newcommand{\kpc}            {\,{\rm kpc}}
\newcommand{\Mpc}            {\,{\rm Mpc}}
\newcommand{\Msun}           {\,{\rm M}_\odot}
\newcommand{\Msunyr}         {\,{\rm M}_\odot\,{\rm yr}^{-1}}

\newcommand{\kmsMpc}         {\,\,{\rm km}\,\,{\rm s}^{-1}\,{\rm Mpc^{-1}}}

\newcommand{\cMpc}            {\,{\rm cMpc}}

\hypersetup{colorlinks = true, linkcolor = green, anchorcolor = red, citecolor = blue, filecolor = red, pagecolor = red, urlcolor = red}

\mathchardef\mhyphen="2D
\defcitealias{Henriques2015}{H15}

\begin{document}

 \volnopage{ {\bf 20XX} Vol.\ {\bf X} No. {\bf XX}, 000--000}
 \setcounter{page}{1}


\title{Merging history of massive galaxies at $3<z<6$}
\author{Kemeng Li\inst{1,2}, Zhen Jiang\inst{3}, Ping He\inst{1,4}, Qi Guo\inst{2}, Jie Wang\inst{2} }
\institute{College of Physics, Jilin University, Changchun 130012, China.
\and National Astronomical Observatories, Chinese Academy of Sciences, Beijing 100012, China; \email{hep@jlu.edu.cn}.\\
\and Department of Astronomy, Tsinghua University, Beijing 100084, China.\\
\and Center for High Energy Physics, Peking University, Beijing 100871, P.R. China.\\ 
 \vs \no
   {\small Received 20XX Month Day; accepted 20XX Month Day}
}

\abstract{The observational data of high redshift galaxies become increasingly abundant, especially since the operation of the James Webb Space Telescope (JWST), which allows us to verify and optimize the galaxy formation model at high redshifts. In this work, we investigate the merging history of massive galaxies at $3 < z < 6$ using a well-developed semi-analytic galaxy formation catalogue. We find that the major merger rate increases with redshift up to 3 and then flattens. The fraction of wet mergers, during which the sum of the cold gas mass is higher than the sum of the stellar mass in two merging galaxies, also increases from $\sim$ 34\% at $z = 0$ to 96\% at $z = 3$. Interestingly, almost all major mergers are wet at $z > 3$ . This can be attributed to the high fraction ($> 50\%$) of cold gas at $z > 3$. In addition, we study some special systems of massive merging galaxies at $3 < z < 6$, including the massive gas-rich major merging systems and extreme dense proto-clusters, and investigate the supermassive black hole-dark matter halo mass relation and dual AGNs. We find that the galaxy formation model reproduces the incidence of those observed massive galaxies, but fails to reproduce the relation between the supermassive black hole mass and the dark matter halo mass at $z \sim 6$. The latter requires more careful estimates of the supermassive black hole masses observationally. Otherwise, it could suggest modifications of the modeling of the supermassive black hole growth at high redshifts.
}

   \authorrunning{Li et al.}            
   \titlerunning{Merging history of massive galaxies}  
   \maketitle

\maketitle

\section{Introduction}           
\label{sect:intro}

The models of galaxy formation and evolution based on the standard cosmological model perform well in explaining the properties of galaxies in the nearby galaxies, however, theories and observations remain unclear in the early universe. With the development of observational facilities in the past decade, several observational results, including massive compact systems up to the first billion years after the Big Bang \citep{miller2018massive,marrone2018galaxy} and the high-z quasars \citep[e.g.,][]{mclure2002measuring,mclure2004cosmological,fan2006evolution,shimasaku2019black}, provide important evidence on the sustained growth mode for massive galaxies in the early universe, especially on their merging history. 

\cite{miller2018massive} reported an extremely concentrated proto-cluster at z $\sim$ 4.3 (SPT2349–56), which contains 14 submillimetre galaxies (SMGs) within a projected region of $130$ kpc, some of the SMGs are likely to undergo a rapid merger and form a massive elliptical galaxy at the core of a cluster with the mass $\sim 10^{15} \Msun$ in the nearby universe. 
\cite{marrone2018galaxy} reported a gas-rich major merger with both galaxy's stellar masses above $10^{10}\Msun$ at $z\sim6.9$ (SPT0311-58) with an extremely high star-formation rate of about a thousand $\Msunyr$. These observational massive compact star-forming galaxies in the early universe contribute more to the star formation rate density and dominate the high-mass galaxy stellar mass function potentially \citep[e.g.,][]{williams2019discovery,forrest2020massive,rennehan2020rapid}. Investigating the incidence of these objects statistically at $z>4$ is meaningful to understand the merging history of massive galaxies at high redshift.

Developing observations of high-redshift quasars imply that supermassive black holes (SMBHs) with $\Mbh \sim 10^{9}\Msun$ formed within the first billion years after the Big Bang. The SMBHs grow rapidly via galaxy mergers predicted by the hierarchical formation of galaxy. However, \cite{shimasaku2019black} claimed that the black hole (BH) mass versus host halo mass relationship at $z\sim6$ indicates a fast growth of black holes at high redshift, as opposed to the milder evolution of the stellar-to-halo mass relation.
Based on the relation proposed by \cite{shimasaku2019black}, \cite{bansal2022evolution} assumed a simplistic model to study how galaxy mergers affect the evolution of SMBHs in the central galaxies, and constrain a lowest mass with $\sim 10^{10}\Msun$ of dark matter halo which can contain a SMBH. It should be interesting to investigate the merging of massive galaxies from the aspect of the black hole-halo mass relation at $z \sim 6$ within the framework of hierarchical structure formation.

The standard model predicts that the formation of these massive systems at high redshifts is related to galaxy mergers, especially major mergers \citep{sanders1990ultraluminous,hopkins2008cosmological,hewlett2017redshift,calabro2019merger}. Galaxy mergers are widely studied in observations and simulations \citep[e.g.][]{lotz2008galaxy,guo2008galaxy,lotz2011major,rodriguez2015merger,qu2017chronicle,mantha2018major} up to $z \sim 3$.
Thus we first examine the ratios and contribution of galaxy merger from $z = 0$ to $6$, with the galaxy catalogue extracted from \cite{guo2011dwarf,guo2013galaxy}. Especially, we attempt to look for observed massive objects in simulation, and investigate the $\Mbh$-$\Mtwoh$ relation as well as the AGN pair fraction statistically. We expect to provide predictions for future observations and constraints for current galaxy formation model. Thus we make use of a large galaxy catalogue extracted from a classical semi-analytic model (SAM) based on the Millennium Simulation (MS) \citep{springel2005simulations}. 

This paper is organized as follows. In Section. \ref{sect:simulation}, we briefly introduce the galaxy catalogue we use in this work. In Section. \ref{sect:results}, we describe the galaxy merger diagram illustrated by our galaxy model and present the incidence of the recent observational massive objects up to $z \sim 6$. We summarize our conclusions in Section. \ref{sect:Conclusion and Disscussion}.

\section{Galaxy Catalogue}
\label{sect:simulation}
In this work, we use the galaxy catalogue (MS-W7) presented in \cite{guo2013galaxy}, which is constructed by implementing the semi-analytic model of \cite{guo2011dwarf} on N-body merger trees extracted from the Millennium Simulation \citep{springel2005simulations} but with WMAP7 cosmology directly.
The Millennium Simulation follows the evolution of $2160^{3}$ dark matter particles from $z\sim127$ to the present day in a comoving box of side length 710 Mpc. The cosmological parameters used in the catalogue are based on WMAP7, with $\Om=0.272$, $\Ob=0.045$, $\Ol=0.728$, $n=0.961$, $\seight=0.807$ and $\Hzero = 70.4 \kmsMpc$. The mass of the simulation particles is $1.3 \times 10^{9} M_{\odot}$, which makes the galaxy catalogue to be nearly complete above stellar mass $10^{9} M_{\odot}$. We refer readers to \cite{guo2011dwarf,guo2013galaxy} for more detailed descriptions on MS-W7.

In this catalogue, we mainly consider galaxies with $M_{*}>10^{10} M_{\odot}$, which is massive enough to resolve the most massive objects we are interested in. 
Several SAMs have been developed in the past decades \citep[etc]{de2007hierarchical,guo2011dwarf,henriques2015galaxy} and achieved good match for the properties of local galaxies, \cite{henriques2015galaxy} even reproduce better stellar mass function at z$\sim$3. However, the predicted galaxy merger rates differ by an order of magnitude in different models \citep{hopkins2010mergers,vulcani2016mergers}. Thus we stick to use the catalogue from \cite{guo2013galaxy}, galaxy merger rate of which matches the observational results well (see more details in fig.\ref{Rm} in section \ref{sect:merg}).

\section{Results}
\label{sect:results}

In Section \ref{sect:merg}, we describe the merger diagram, especially major mergers, across redshift 0-6 for galaxies with $\Mstar > 10^{10} \Msun$. In Section \ref{sect:massive}, we investigate the incidence of of several certain observed massive objects, including massive gas-rich galaxy pair, proto-clusters and SMBHs, and the fraction of AGN pairs in the merging black holes.

\subsection{Ratios and contributions of mergers up to redshift 6}
\label{sect:merg}
In this section, we investigate the evolution of galaxy assembly through galaxy mergers up to $z\sim6$. In semi-analytic models, a galaxy merger takes place when a galaxy has more than one direct progenitor. Direct progenitors of a galaxy could be found in adjacent or nonadjacent previous snapshots. If a galaxy has m ($m\ge2$) direct progenitors, the galaxy is assumed to have experienced $m-1$ mergers between these two snapshots. Merger events can be classified into different types according to their stellar mass ratio of the two merging galaxies: $\mu \equiv \Mstars/\Mstarp (\Mstarp \ge \Mstars)$, where $\Mstarp$ and $\Mstars$ represent the stellar mass of the primary and secondary galaxy, respectively. A merger is defined as major merger if $\mu \geq 1/4$; minor merger if $1/10 \leq \mu \leq 1/4$; and accretion if $\mu < 1/10$, which is not considered in this work. We also classify mergers according to their gas to stellar mass ratio: $\fgaspair \equiv (\Mgasp+\Mgass)/(\Mstarp+\Mstars)$. Agreed with previous works \citep[e.g. ][]{lagos2018quantifying}, a merger is defined as a wet (gas-rich) merger if $\fgaspair > 0.5$, otherwise it is a dry (gas-poor) merger. Note that in the semi-analytic model we use in this work, ``cold gas'' indicates atomic and molecular gas which composes the gaseous disk.

\begin{figure}
    \centering
    \includegraphics[width=\columnwidth]{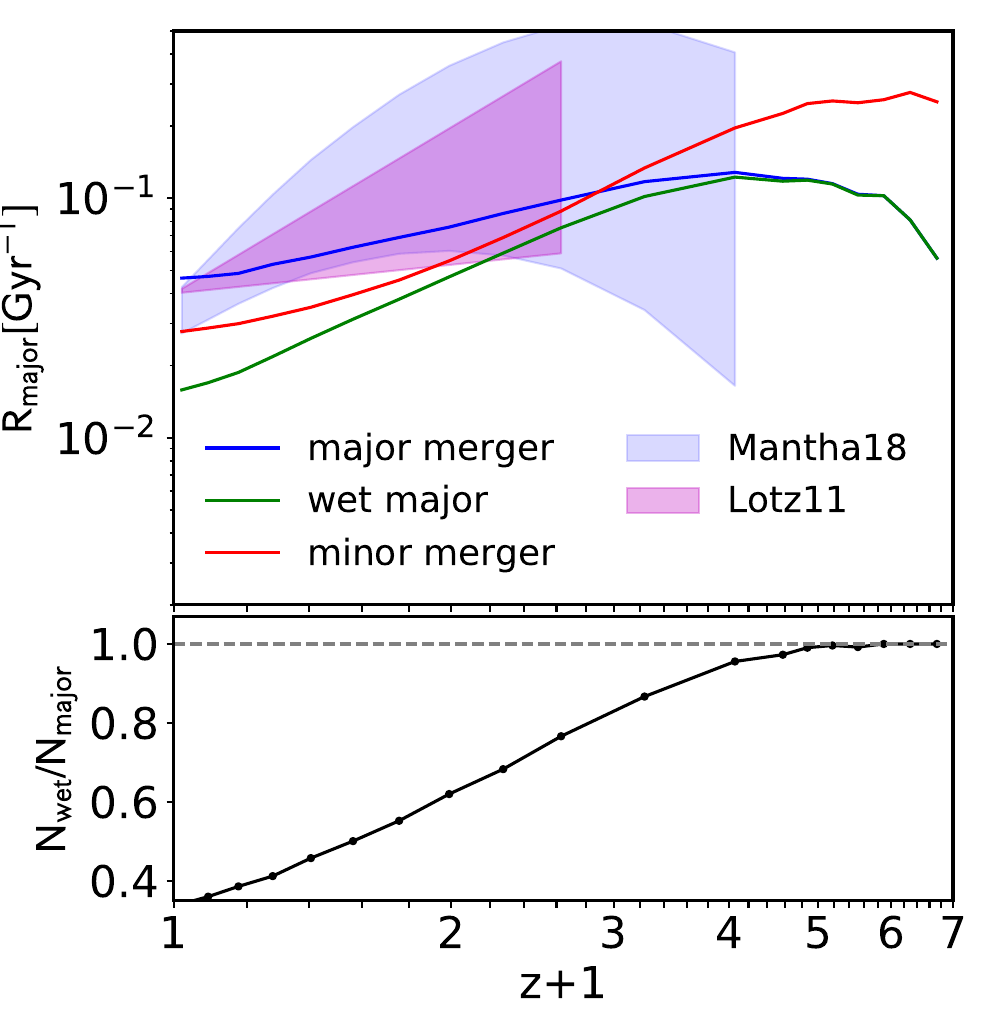}
    \caption{Upper panel: The major merger rate (blue solid line), wet major merger rate (green solid line) and minor merger rate (red solid line) of galaxies with $\Mstar > 10^{10}\Msun$ as a function of redshift. The magenta and purple shaded regions show the observational constraints of \protect\cite{lotz2011major} and \protect\cite{mantha2018major}, respectively. Lower panel: The fraction of wet major mergers in all major mergers. The horizontal grey dashed line represents unity.}
    \label{Rm}
\end{figure}

We describe the galaxy merger rate as $\Rm=\fmerg/\tau$, where $\fmerg$ is the galaxy merger fraction, calculated by dividing the number of galaxies involved in merging events by the total number of galaxies at each snapshot, and $\tau$ is the time interval between the snapshots of the descendant galaxies and their direct progenitors in the simulation. In the upper panel of Fig.~\ref{Rm}, we show the galaxy merger rate whose stellar mass is above $10^{10} \Msun$ as a function of redshift over $z \sim 0-6$. We can see that the minor merger rate increases with redshift monotonically across $z\sim0-6$. The major merger rate also increases with redshift at lower redshift but levels off at $z\sim3$. At $z\lesssim2$, the minor merger rate is lower than the major merger rate slightly, and becomes higher than the major merger rate by about a factor of 2 up to $z>3$. A higher frequency of the major merger with the same definition of major and minor merger within a similar mass range is similar to the result of Emerge \citep{o2021emerge}. Also, the merger rate is related to the upper and lower limit of the stellar mass ratios \citep{guo2008galaxy}, as well as the infalling efficiency. The major merger rate should be lower than the minor merger rate if we choose the progenitor stellar mass ratio of 3 to 1.

Within the observed redshift range of \cite{lotz2011major} at $z\leq1.5$ the magenta region) and \cite{mantha2018major} at $z\le3$ (the purple region), our major merger rate agrees with the observational constraints quite well, though the uncertainties in observations are relatively large. Note that the galaxies used in \cite{mantha2018major} are more massive than $2 \times 10^{10} \Msun$. We divide their results by a factor of $1.30-1.45$ at each redshift to match the samples in this work, as the major merger rate of galaxies with $\Mstar >2 \times 10^{10} \Msun $ is $1.30-1.45$ times the one with $\Mstar > 10^{10} \Msun $ at each redshift in our catalogue.

\begin{figure*}
    \centering
    \includegraphics[width=2\columnwidth]{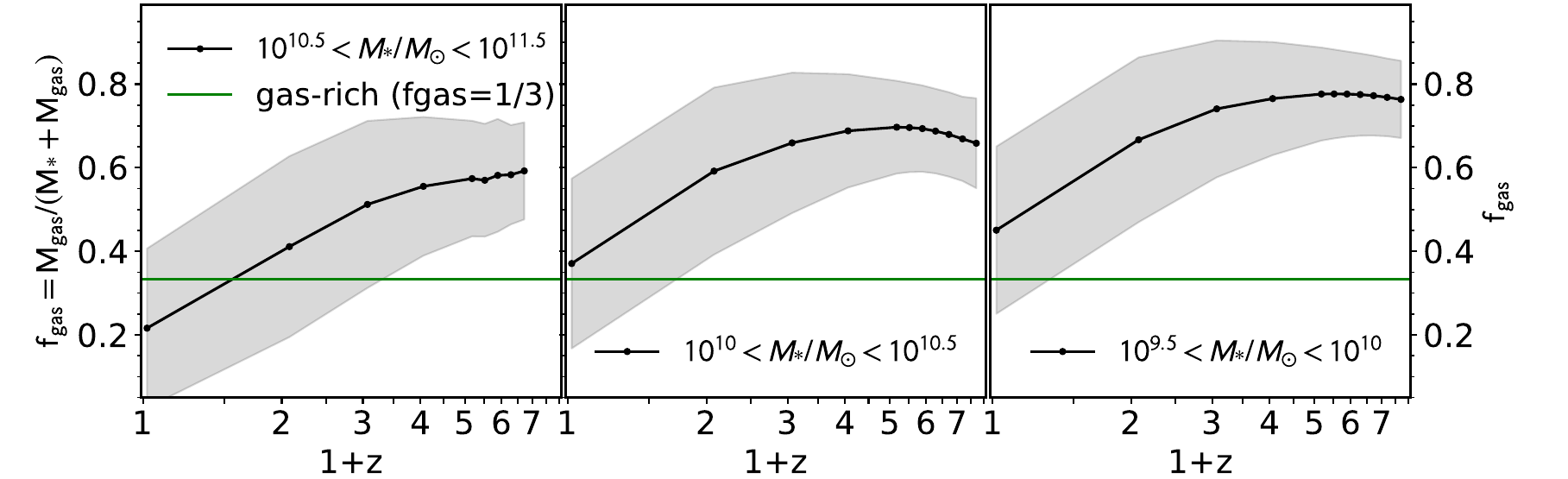}
    \caption{The mean cold gas fraction as a function of redshift for the three galaxy classes in the stellar mass bin of $10^{10.5}-10^{11.5}\Msun$ (left), $10^{10}-10^{10.5}\Msun$ (center), and $10^{9.5}-10^{10}\Msun$ (right). Shaded regions present 1$\sigma$ scatter of the distributions. The green line presents $f_{gas}\equiv\Mgas/(\Mstar+\Mgas)= 1/3$, which is the lower-limit for the so-called gas-rich galaxies. If both merging galaxies are gas-rich, the merger will be a wet merger.}
    \label{MgMs}
\end{figure*}

In the lower panel of Fig.~\ref{Rm}, we show the evolution of wet major merger fraction. We can see that the gas-rich fraction of major mergers increases from $34\%$ to nearly $100\%$ from redshift 0 to 3, which indicates that wet major mergers play a crucial role in the assembly of galaxies at high redshifts. This increase is a natural result of the increasing fraction of cold gas in merging galaxies. Fig.~\ref{MgMs} shows the evolution of the cold gas fraction of galaxies, defined as $\fgas\equiv\Mgas/(\Mstar+\Mgas)$, in three different mass ranges. As shown in Fig.~\ref{MgMs}, the gas fraction increases by about two times from redshift 0 to 4 and low-mass galaxies have more cold gas compared with massive ones. We notice that even in the most massive range, with $\Mstar=10^{10.5}-10^{11}\Msun$, almost all galaxies have $\fgas>1/3$ above $z\sim3$, represented by the green horizontal line in Fig.~\ref{MgMs}. If both merging galaxies have $\fgas>1/3$, we have $\fgaspair>0.5$, then it will be a wet merger according to our definition. Thus the evolution of the gas fraction in galaxies is consistent with the evolution of the gas-rich fraction in major mergers.

To evaluate the relative importance of different merger types to galaxy growth, we calculate fractional mass contributions ($\fmass$) of different merging channels, including major mergers, wet major mergers, and minor mergers. The fractional mass contribution is defined as the merged stellar masses fraction of the final stellar mass across the assembly history, under the assumption that all the mass of these objects is accreted by the main progenitor, i.e. the most massive progenitor. Note that in this work, we only focus on central galaxies, avoiding the effects of some environmental processes such as strangulation and ram-pressure stripping. In \cite{guo2011dwarf}, satellites that have lost their host subhaloes could be totally disrupted to be intracluster stars by very strong tidal forces, depending on orbital parameters and mass ratio. We neglect contributions of these disrupted satellites.

\begin{figure}
	\centering
    \includegraphics[width=\columnwidth]{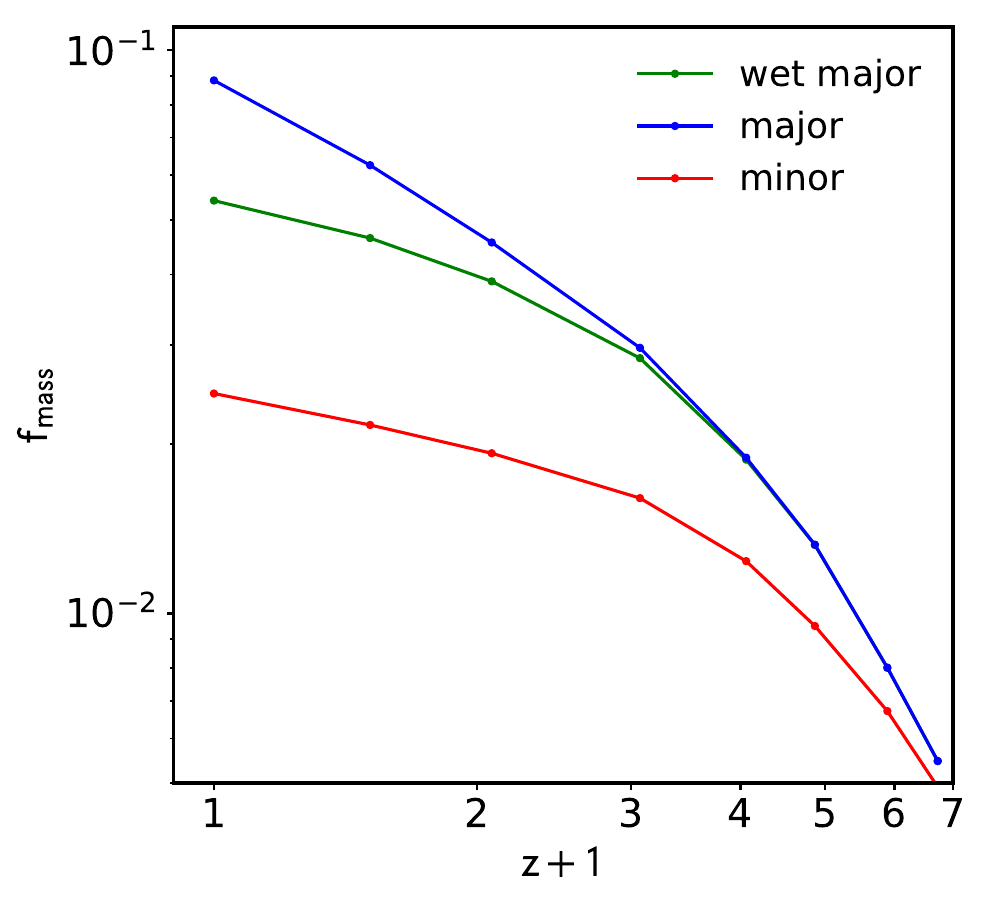}
    \caption{The mean fractional stellar mass contributions of major mergers (blue line), wet major mergers (green line) and minor mergers (red line) for galaxies with $\Mstar>10^{10}\Msun$ at different redshifts. As the assembly histories of galaxies are quite different, the scatter of $\fmass$ is huge (see Fig.~\ref{f_expN}). Here we only plot the mean relation for simplicity.}
    \label{MassC}
\end{figure}

The fractional mass contributions of different merger types are shown in Fig.~\ref{MassC}. We can see that the mean contribution of major mergers decreases from $8.84\%$ to $0.55\%$ from redshift 0 to 6, and that of minor mergers decreases from $2.26\%$ to $0.49\%$. We find that there is only $\sim10\%$ of the final stellar mass is contributed from major and minor mergers for local galaxies with $\Mstar>10^{10}\Msun$. Limited by the resolution of our simulation, we don't consider the contribution from accretion. Nevertheless, \cite{qu2017chronicle} found that the fractional mass contribution from external processes, including mergers with all mass ratios, is $10\%-20\%$ from redshift 0 to 3, and major mergers make more contributions in the more massive galaxies. Moreover, major mergers make a greater contribution to galaxy assembly than minor mergers across redshift $0-6$, even though the major merger rate is lower than the minor merger rate at $z \gtrsim 2$. 

\begin{figure}
	\centering
    \includegraphics[width=\columnwidth]{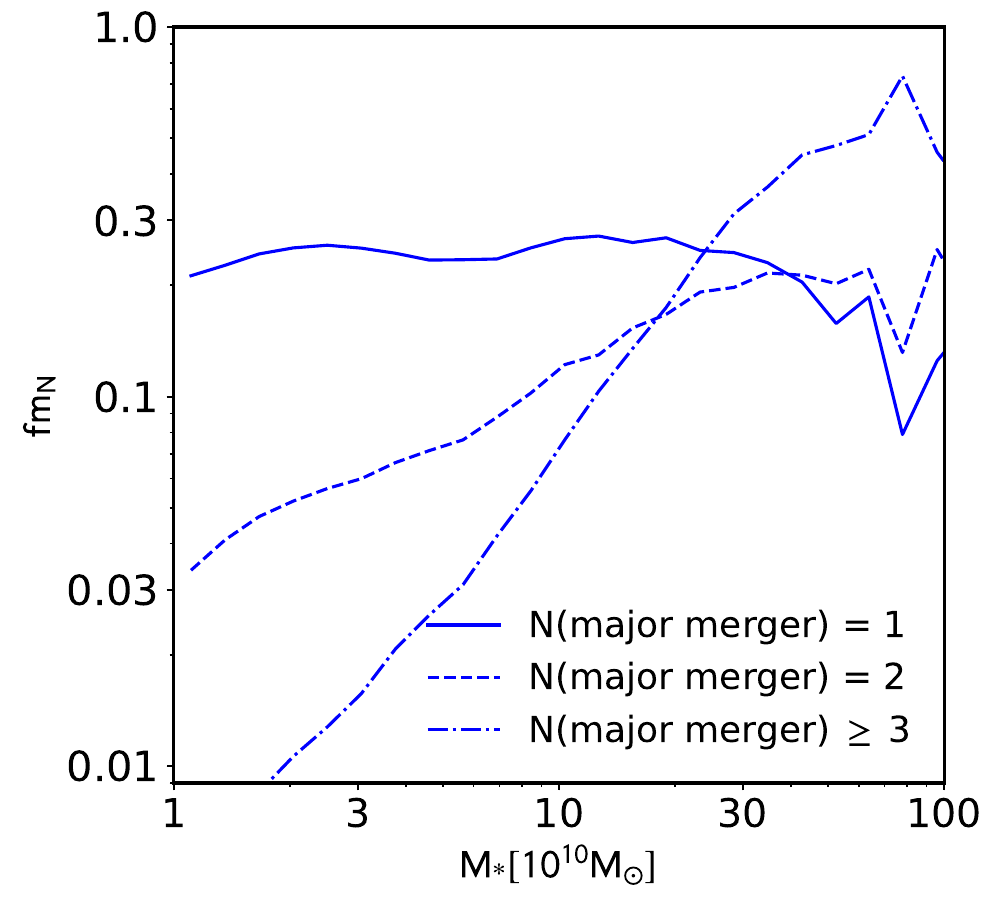}
    \caption{The number fraction of local galaxies that have experienced one major merger (solid line), two major mergers (dashed line) and at least three major mergers (dot-dashed line) till $z\sim0$ as a function of stellar mass.}
    \label{f_expN}
\end{figure}

To illustrate the role of major mergers in galaxy assembly, we then investigate how many local galaxies have experienced more than one major merger through their formation history. The number fraction of the galaxies that have experienced $N$ times of major mergers ($\fmN$) till $z\sim0$ is calculated in Fig.~\ref{f_expN}. We find that $fm_{\rm N=1}$ is constantly about $26\%$ over stellar mass range $10^{10}-10^{11}\Msun$ and decreases slightly in the most massive end. $fm_{\rm N=2}$ increases mildly with stellar mass. It increases by a factor of 2 from the stellar mass of about $10^{10}\Msun$ to the most massive end. $fm_{\rm N\ge 3}$ increases rapidly with stellar mass and reaches $50\%$ up to $\Mstar\sim3\times10^{11}\Msun$. It’s clear that more massive galaxies have experienced more major mergers throughout their assembly history. Therefore, detailed studies on the major mergers are essential to understand the evolution of most massive galaxies.

\subsection{Observational massive objects at high redshifts}
\label{sect:massive}
In this section, we compare the simulation data with observational measurements of massive gas-rich pair, and distribution ranges of massive proto-clusters, central massive black hole masses, as well as DAGNs. We investigate counterparts of observed massive objects at $z>3$ in the galaxy catalogue, which should contribute to understand the hierarchical formation in the distant universe.

\subsubsection{Massive gas-rich galaxy pair up to redshift 6}
\begin{figure}
	\centering
    \includegraphics[width=\columnwidth]{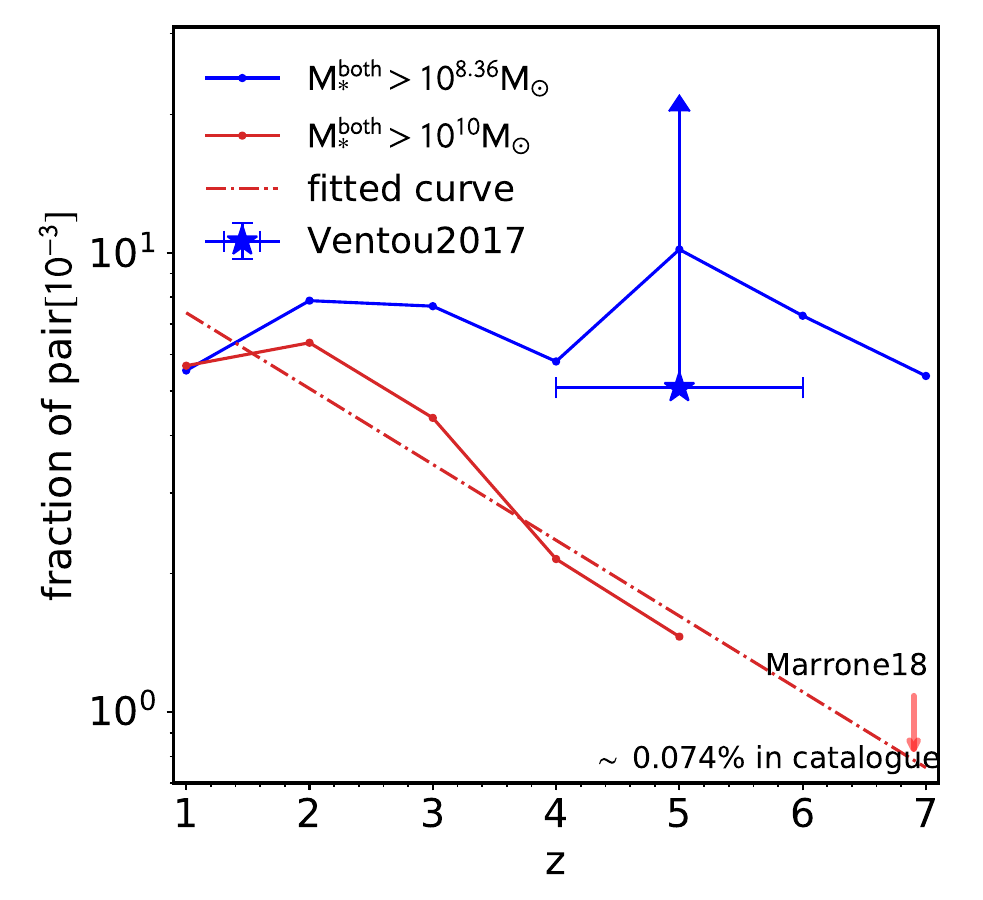}
    \caption{The gas-rich major merger fraction with both stellar masses ($\Mboth$) above a given stellar mass as a function of redshift. The blue and red solid lines represent the fraction of wet major mergers with both stellar masses above $10^{8.36}\Msun$ and $10^{10}\Msun$ in the simulation, respectively. The blue star with arrow indicates the observational constrained lower limit for $\Mboth > 10^{8.36}\Msun$ of \protect\cite{ventou2017muse}. The red dot-dashed line fits the evolution of the fraction with $\Mboth > 10^{10}\Msun$ at redshift $1-5$ in our galaxy catalogue. The red arrow marks the corresponding redshift of the object in \protect\cite{marrone2018galaxy}.}
    \label{fm_z}
\end{figure}

Motivated by a gas-rich major merger with both stellar masses above $\sim 10^{10}\Msun$ at $z \sim 6.9$, observed through far-infrared in \cite{marrone2018galaxy}, we explore the fraction of such gas-rich major mergers with both stellar masses ($\Mboth$) above a certain value at high redshift. First, we compare the gas-rich major merger fraction of galaxies with $\Mboth > 10^{8.36}\Msun$ with data from the MUSE Hubble Ultra Deep Field Survey \citep{ventou2017muse} \footnote{Note that almost all major mergers are gas-rich at $z>3$ in our simulation, as shown in the lower panel of Fig.~\ref{Rm}, allowing us to compare the gas-rich major merger fraction in simulation with the spectroscopic close pair counts in observation directly.}, which covers a $1 \times 1 arcmin^{2}$ area, wavelegth $4750-9300\overset{\circ}{A}$, as shown by the blue lines in Fig.~\ref{fm_z}. The pair fraction of galaxies with $\Mboth > 10^{8.36}\Msun$ in our galaxy catalogue is about 0.7\% across the redshift range $1-7$ and agrees well with the observational constraints. That is to say, on average we could find one major merger with $\Mboth > 10^{8.36}\Msun$ among 142 galaxies at redshift $4-6$ without any other selection effect. We then explore how many cases described in \cite{marrone2018galaxy} could be found in our galaxy catalogue. As shown by the solid red line in Fig.~\ref{fm_z}, the major merger fraction with $\Mboth > 10^{10}\Msun$ decreases from 0.6\% to 0.1\% from redshift 1 to 5. It's number density is about $10^{-8} \cMpc^{-3}$ at $z\sim5$. Nevertheless, we find that there is no such massive pair at $z>5$ in our galaxy catalogue. We fit the evolution of the fraction at redshift $1-5$ and extrapolate it to higher redshifts, as indicated by the red dot-dashed line. We predict that on average the pair fraction with $\Mboth > 10^{10}\Msun$ at $z\sim7$ is about one over 1430 galaxies, which is one magnitude lower than the one with $\Mboth>10^{8.36}\Msun$.
Note that the galaxy catalogue we used in this work shows a deficit at the massive end of stellar mass function above redshift 2, especially for galaxies with $\Mstar > 10^{10.5} \Msun$ at $z>3$ \citep[see ][]{guo2011dwarf, guo2013galaxy}, which could be responsible for the lack of such massive pairs at $z>5$.

\subsubsection{Proto-cluster up to redshift 4}
\cite{miller2018massive} reported a uniquely dense massive system at $z \sim 4.3$, which consists of at least 14 gas-rich star-forming sub-millimetre galaxies (SMGs), whereas they are located within a projected region that is only around $130\kpc$ in diameter. These 14 SMGs indicate the dynamical halo mass to be $10^{13}\Msun$. Some other observed proto-clusters that contain as many SMGs as this system extends over much larger areas in the sky. Such a dense system is likely to merge to form a massive elliptical galaxy at the core of a galaxy cluster in the nearby universe.

\begin{figure}
	\centering
    \includegraphics[width=\columnwidth]{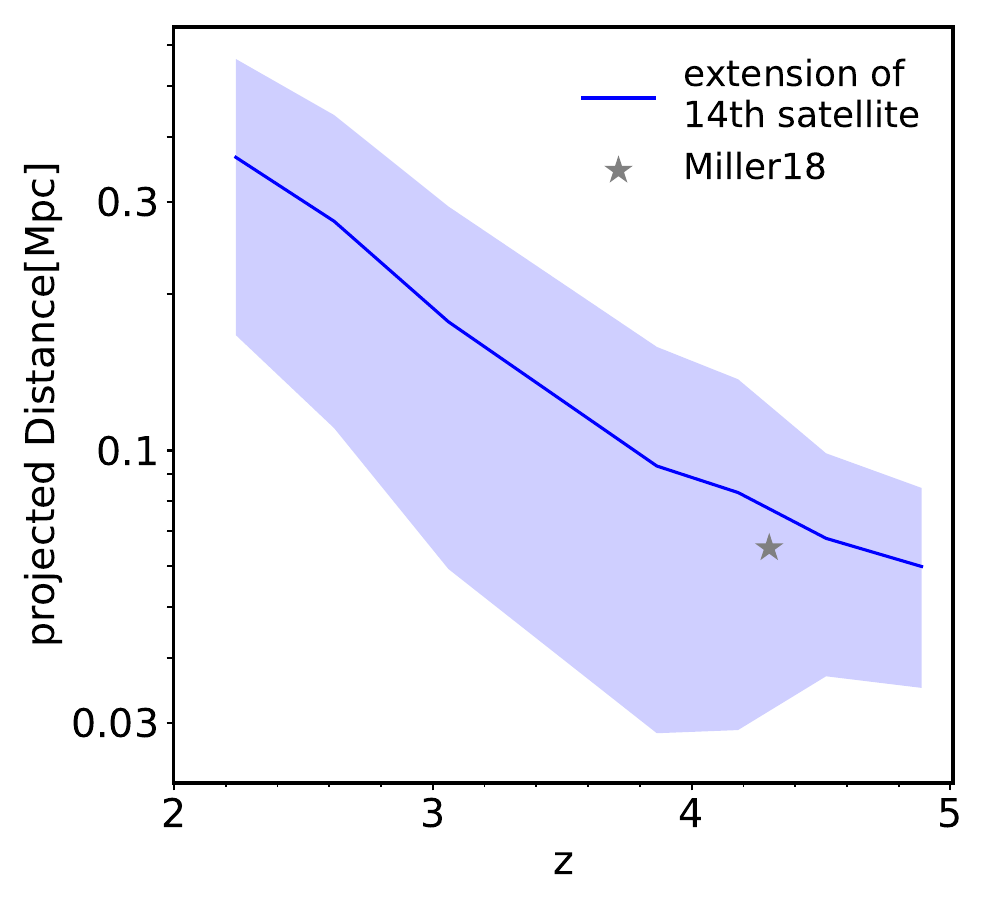}
    \caption{The extension of gas-rich ($\fgas>1/3$) star-forming (star formation rate $>10\Msunyr$) satellite systems within massive haloes with $10^{12.8} \leq \Mtwoh/\Msun \leq 10^{13.5}$ at each redshift in our galaxy formation model, i.e. the projected physical distance from the 14th gas-rich star-forming satellite to the central galaxy. The coloured region shows corresponding $1\sigma$ scatter. The grey star represents the extension of the observed system of 14 SMGs (the radius of the projected region, $0.065\Mpc$) at $z\sim4.3$ in \protect\cite{miller2018massive}.}
    \label{d_N}
\end{figure}

To explore the incidence of such a dense system, we investigate how gas-rich ($\fgas>0.3$) star-forming (star formation rate $>10\Msunyr$) satellites ($\Mstar\geq10^{9}\Msun$) distribute within massive haloes with $10^{12.8} \leq \Mtwoh/\Msun \leq 10^{13.5}$ over redshift $2-5$, where $\Mtwoh$ is halo mass, the total mass of particles within the radius where a mean overdensity exceeding 200 times the critical value. We evaluate the extension of a satellite system as the physical distance from the $N$-th satellite to the central galaxy. Especially, we use projected distance from a random line of sight to compare with that in the observation. As shown in Fig.~\ref{d_N}, the extension of satellite systems consisting of 14 gas-rich star-forming satellites are $0.41\Mpc$ and $0.06\Mpc$ at redshift 2 and 5, respectively. We find that the system reported in \cite{miller2018massive} is well located within the 1$\sigma$ scatter of the distribution, indicating such dense massive system is not rare at high redshift in our galaxy formation model. The projected distance is only about $1/2-1/3$ of the physical distance, which implies that we need to be careful when we deal with projected distances. The effect of projection is a considerable factor if we attempt to compare the extension of an observed satellite system.

\subsubsection{Super massive black holes and active galactic nucleus}
\cite{shimasaku2019black} reported the relation between the mass of super massive black holes (SMBHs) and the mass of their host dark matter haloes for 49 quasi-stellar objects (QSOs) at $z\sim6$. They found the vast majority of these $z\sim6$ SMBHs are more massive than expected from the local $\Mbh$–$\Mtwoh$ relation. While the evolution of the stellar-to-halo mass ratio is much milder, especially for haloes with $\Mtwoh < 10^{12.5} \Msun$ \citep{girelli2020stellar}, these results imply a rapid SMBH growth in dark matter haloes at $z\ge6$.

\begin{figure}
	\centering
    \includegraphics[width=\columnwidth]{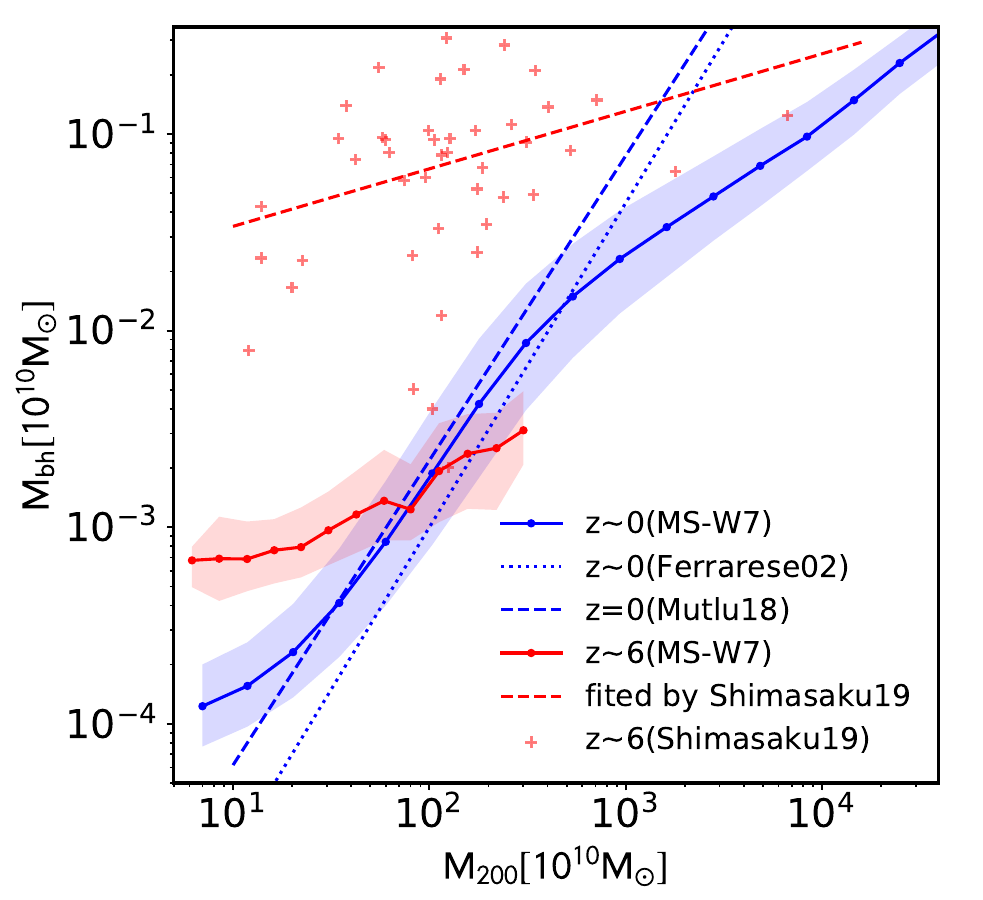}
    \caption{BH masses as a function of their host halo masses. Blue and red solid lines represent the relation in our galaxy catalogue at redshift 0 and 6, respectively, 
    and the coloured regions indicate corresponding 25th and 75th percentiles of the distributions. 
    The red pluses are the observational results at $z\sim6$ of \protect\cite{shimasaku2019black}, and the red dashed line is their fitting. The blue dashed line is fitted by local observational results of \cite{ferrarese2002beyond}, and the blue dotted line is extracted from Illustris simulation by \cite{mutlu2018illustris} at $z\sim0$.}
    \label{MbhMh}
\end{figure}

We present $\Mbh$ versus $\Mtwoh$ relation at $z\sim0$ and $6$ in the galaxy formation model in Fig.~\ref{MbhMh}, as shown by solid lines with different colours. Our results share a similar trend with previous works \citep[e.g.,][]{ferrarese2002beyond,mutlu2018illustris}. However, the slope of local relation in our galaxy catalogue is shallower than the best-fitted relation in observation \citep{ferrarese2002beyond} and the prediction of Illustris hydrodynamical simulation \citep{mutlu2018illustris}. This deviation could be caused by the estimation of halo mass which depends on the relationship of viral velocity, circular velocity and rotation velocity in observation, or by the different BH growth mechanisms in galaxy models. We notice that most BH masses of the observed QSOs in \cite{shimasaku2019black} are above $\sim 10^{8}\Msun$ even in haloes with $\Mtwoh = 10^{11} - 10^{12}\Msun$ at $z\sim6$, and the most massive ones are even above $10^{9}\Msun$, about two magnitudes more massive than our model predictions. Such a deviation from local relation is perhaps caused by the lack of low-luminosity BHs in observations at high redshift. But unfortunately, the BH growth mechanism mainly driven by major mergers in our galaxy formation model fails to reproduce these BHs at $z\sim6$, which implies that more physical details are needed about BH growth.

\begin{figure}
	\centering
    \includegraphics[width=\columnwidth]{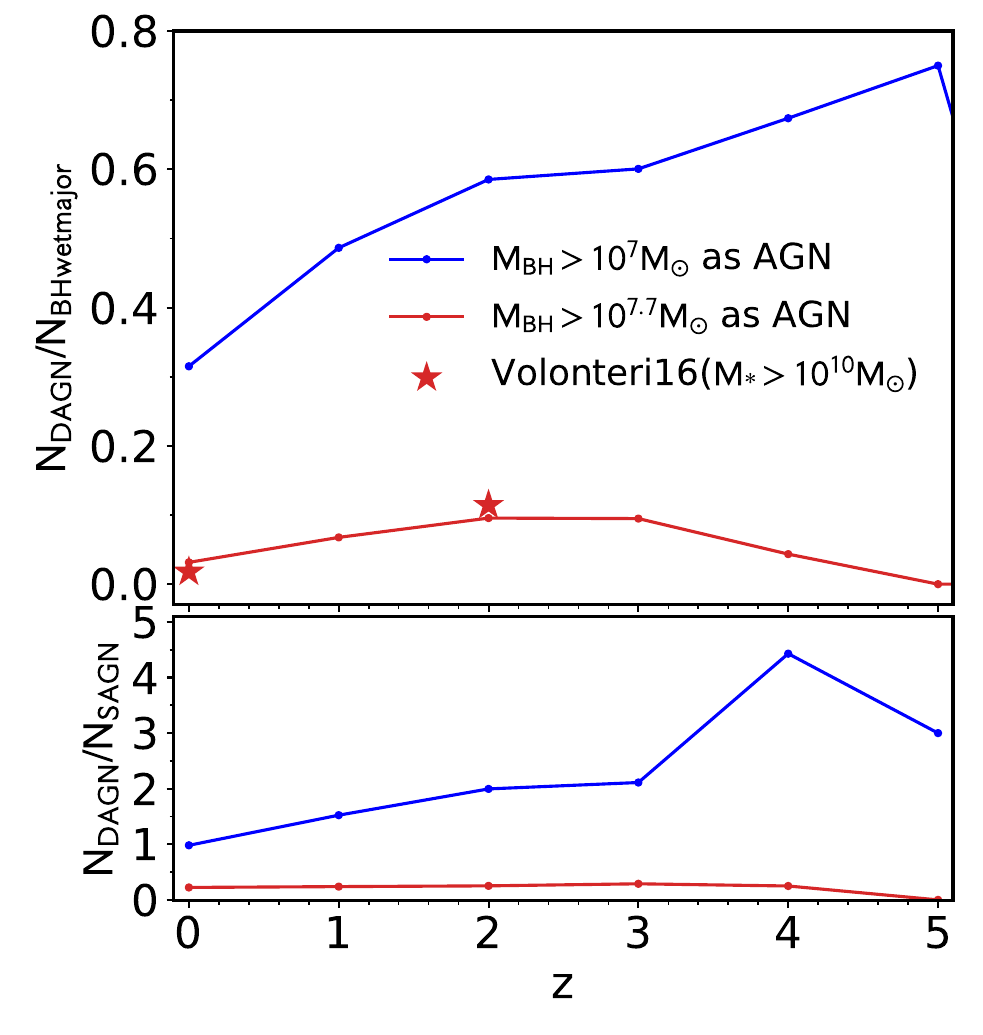}
    \caption{Upper panel: Dual AGN (DAGN) fraction of primary galaxies with stellar masses above $10^{10}\Msun$. Blue and red lines indicate results with different BH critical mass of AGN in wet BH major mergers (blue: $\Mcrit=10^{7}\Msun$; red: $\Mcrit=10^{7.7}\Msun$). Red stars mark the hydrodynamical simulation result of \cite{volonteri2016cosmic}. Lower panel: The number ratio of DAGNs to single AGNs (SAGNs) in gas-rich black hole mergers.}
    \label{Mstar_QSO}
\end{figure}

As a census of dual AGNs (DAGNs) across cosmic history is important to study BH binaries and reconcile galaxy mergers with BH growth, and major mergers are believed to trigger AGNs at low redshift, we attempt to explore the evolution of DAGN fraction ($\fDAGN$) in gas-rich BH major mergers with primary stellar mass above $10^{10}\Msun$ up to redshift 5. In wet mergers, we considerdefine a BH as an AGN if its mass is larger than a critical value $\Mcrit$. If both merged BHs meet this criterion, it is considered as a DAGN system, otherwise it is a single AGN (SAGN) or not a AGN system. Since observations show that BH masses of most quasars have the lowest limit of $10^{7}\Msun$ \citep{mclure2004cosmological}, we first consider a critical mass $\Mcrit=10^{7}\Msun$. As shown in the top panel of Fig.~\ref{Mstar_QSO}, $\fDAGN$ with $\Mcrit=10^{7}\Msun$ increases with redshift across redshift $0-5$. Moreover, we find that a critical mass $\Mcrit=10^{7.7}\Msun$ matches the hydrodynamical simulation result of \cite{volonteri2016cosmic} \footnote{\cite{volonteri2016cosmic} presents the DAGN fraction in major mergers. Here we divide their result by the wet merger fraction in major mergers at $z=0,2$ to get the DAGN fraction in wet major mergers.} well, as shown by the red line. $\fDAGN$ with $\Mcrit=10^{7.7}\Msun$ increases with redshift until it reaches the peak at $z\sim3$. In the lower panel of Fig.~\ref{Mstar_QSO}, we also present the number ratio of DAGNs to SAGNs in gas-rich BH mergers.
We find that the ratio of DAGNs to SAGNs defined by the lower mass cut increases from 1.0 to 2.1, which indicates that there are more DAGNs than SAGNs, if low mass black hole is bright enough at high redshift. While with the higher mass cut definition, DAGNs are rare (only $\sim25\%$ of SAGNs) across the cosmic time.

\par

\section{Conclusion and Discussion}
\label{sect:Conclusion and Disscussion}
In this work, we use the galaxy catalogue of \cite{guo2011dwarf,guo2013galaxy} in the Millennium simulation to explore galaxy-galaxy merger, especially major mergers across redshift $0-6$. The major merger rate matches observation constraints well up to $z\sim3$. Moreover, we find that almost all major mergers are gas-rich at $z>3$ which is the result of the increasing gas fraction in galaxies with increasing redshifts.

Because the high ratio of the wet major mergers at $z>3$, to understand the starburst, BH rapid growth as well as AGN active during wet major mergers are crucial to the modeling of galaxy formation at high redshifts. We believed that the resolved cold gas phases (HI and HII in disks) in semi-analytic models developed by \cite[][etc]{Xie2017,henriques2020galaxies}, which could be used to compare with future high-redshift observations \citep{sun2020probing}, should help to refine the processes of star forming during wet mergers. Furthermore, the galaxy merger diagram in our model, including the merger rate and the effect of major merger on the transition of galaxy and black hole evolution in the early universe, should be checked carefully combined with future observations \citep[e.g.,][and Muzzin et al in prep]{faisst2019alma,gonzalez2019atacama,marsan2022number}.

Furthermore, we compare our model predictions with some merging phenomena on massive galaxies $3 < z < 6$, and some some main conclusions as the below individually:

\begin{enumerate}

\item \textbf{The incidence of gas-rich major merger:} We predict that on average one wet major merger case with $\Mboth  > 10^{10}\Msun$ can be found among $\sim 1430$ galaxies with comparable mass at $z\sim7$, though the selection criterion of dust and gas fraction is loose relative to the observed systems in \cite{marrone2018galaxy}. Also, we notice that our catalogue shows a lack of massive galaxies at $z>3$, one of improvements on this issue is addition constraints on SN feedback \citep[e.g.,][]{henriques2015galaxy,Hirschmann16,Fontanot_2017}.

\item \textbf{The incidence of the dense starburst core in a cluster:} We found a dense system consisted of 14 gas-rich star-forming satellites could locate in an average projected physical region with a radius $\sim 0.06$ Mpc in a cluster at $4<z<5$, while the similar system found in \cite{miller2018massive} locates within the $1\sigma$ scatter of the distribution of datas in simulation. This indicates such dense massive system is not rare at high redshift underlying our galaxy formation model. Also, improving the precision of measurement in the high-z observation is important in the future exploration.

\item \textbf{The black hole-halo mass relation:} In our catalogue, although the relationship between SMBH mass and bulge stellar mass at $z\sim0$ satisfies the observed scaling relation, the local relationship between SMBH mass and host halo mass deviates from observational constraints slightly. Besides, our galaxy model fails to reproduce SMBHs with $\Mbh > 10^{8} \Msun$ at $z\sim6$. This suggests that the BH growth in early universe needs additional mechanisms.

\item \textbf{The dual AGN fraction in wet major merger:} AGNs are simply defined as BHs above a critical mass in wet major mergers. If we assume a lowest critical BH mass of $10^{7}\Msun$, the DAGNs fraction increases with redshift and reaches the peak at $z \sim 5$, and the ratio of DAGNs over single AGNs (SAGNs) increases from 1 to 2 times over redshift $0-3$. Whereas, if we assume a more realistic critical mass of $10^{7.7}\Msun$, the DAGNs fraction reaches the peak at $z \sim 2$ and decreases to zero at $z \sim 5$.

\end{enumerate} 

Thus our galaxy model could reproduce the observational massive merging events at high redshift, while fails to reproduce the black hole-halo mass relation and the super massive black holes at $z \sim 6$ as we have observed in a decade. As the AGN feedback is a key ingredient of galaxy evolution in simulations, it should be interesting to explore how to seed a black hole in early universe in SAMs, and to check the reliability of the black hole seeding and growth model from different simulations. Combining with that more optically obscured AGNs are discovered by Wide-Field Infrared Sky Explorer high-sensitivity hard-X-ray observations \citep[e.g.,][]{satyapal2014galaxy,koss2018population} within the BH separation of several kpc, we believe that the accretion mode of BHs in mergers, as well as the galaxy formation model in the early universe should be optimised in the future. Also, with the operation of JWST, more observational data on the merging history of massive galaxies at high redshift will be available, we will be in a golden age to verify the formation history of galaxies during their earlier lifes. 

\normalem
\begin{acknowledgements}
This work is supported by the National Key R\&D Program of China (No.2018YFE0202900 and 2022YFA1602901), the NSFC grant (No 11988101, 11873051, 12125302), CAS Project for Young Scientists in Basic Research Grant (No. YSBR-062), and the K.C.Wong Education Foundation. Q.G acknowledges the National Key Research and Development of China (No.2018YFA0404503), NSFC grants
(No.12033008), the K.C.Wong Education
Foundation, and the science research grants from the
China Manned Space Project with NO.CMS-CSST-2021-
A03 and NO.CMS-CSST-2021-A07.
PH acknowledges the support by the National Science Foundation of China (No.12047569, 12147217), and by the Natural Science Foundation of Jilin Province, China (No. 20180101228JC).

\end{acknowledgements}
  
\bibliographystyle{raa}
\bibliography{a_merger}

\end{document}